\documentclass{PoS}

\def\lsim{\raise0.3ex\hbox{\,$<$\kern-0.75em\raise-1.1ex\hbox{$\sim$}\,}}
\def\gsim{\raise0.3ex\hbox{\,$>$\kern-0.75em\raise-1.1ex\hbox{$\sim$}\,}}

\title{Electron in a transverse harmonic potential}

\ShortTitle{Electron in a transverse harmonic potential}

 \author{\speaker{H. Honkanen}%
        \thanks{This work was done in collaboration with P.~Maris, J.~P.~Vary and S.~J.~Brodsky, and was supported in part by a DOE Grant DE-FG02-87ER40371.
Computational resources were provided by DOE through the
National Energy Research Supercomputer Center (NERSC).}\\
        Iowa State University, USA\\
        E-mail: \email{heli@iastate.edu}}

\abstract{
Non-perturbative solutions to the quantum-field theory is a topic of current 
and broad interest, especially for the heavy ion and laser physics
communities, since they investigate particle production in
the presence of strong external fields. We have
solved a non-perturbative
QED + external field problem of an electron in a strong transverse confining 
potential using Hamiltonian light-front quantum field theory  in
a basis function representation.
The invariant mass 
spectra and the anomalous magnetic moment of the lowest state for this
two-scale system are also evaluated. With this method the perturbative QED
results are also reproduced with a good accuracy. We also discuss the 
extension of the method to other problems as well. }

\FullConference{Light Cone 2010 - LC2010\\
		June 14-18, 2010\\
		Valencia, Spain}

\begin{document}
\section{Motivation}\label{motivation}
The Hamiltonian light-front formalism is ideal for solving problems
non-perturbatively,  since time is
set along the light-front and evaluated experimental observbles, such as 
masses,  form factors, and structure function, are Lorentz frame independent.
Light-front Hamiltonian quantum field theory also has 
similarities with non-relativistic quantum many-body theory. This 
connection was exploited in \cite{Vary:2009gt}, where
 a ``Basis Light Front Quantized (BLFQ)" approach was outlined
by adopting a light-front single-particle basis space consisting of the 2-D 
harmonic oscillator for the transverse modes and a discretized momentum space 
basis for the longitudinal modes.  Adoption of this basis is also consistent 
with recent developments 
in AdS/CFT correspondence with QCD \cite{deTeramond:2008ht}.

In \cite{Honkanen:2010rc} this approach was applied to
address the problem of an electron in a transverse harmonic cavity with
the QED Hamiltonian evaluated on the light-front in a 
Fock space consisting of electron states and  electron plus photon states.
The external field was included non-perturbatively and 
the eigenvalues, eigenvectors and anomalous magnetic moment were solved.
In a non-renormalized case, the obtained electron anomalous magnetic moment 
was within
1.5\%  of the theoretically expected result (Schwinger moment), when 
extrapolated to the
zero external field limit. Applying a sector-dependent renormalization scheme 
\cite{Karmanov:2008br} to this Hamiltonian, the zero external field results 
were consistent 
with related works  Refs.\cite{Brodsky:1998hs,Brodsky:2004cx}  and 
Refs.\cite{Chabysheva:2009ez,Chabysheva:2009vm},
where the one-photon truncated
light-front Hamiltonian was regulated with a Pauli-Villars regularization 
scheme.

The nonperturbative analysis presented in \cite{Honkanen:2010rc}  could be 
applicable
to measurements of the (gyromagnetic) ratio of the spin precession to Larmor 
frequencies of a trapped electron in strong external electromagnetic fields,
and can be straightforwardly extended by 
incorporating higher Fock-space sectors.
It also serves as a first step towards the studies of non-perturbative QED
relevant for the anomalous enhancement of lepton production
at RHIC \cite{Adare:2009qk} and for proposals for producing
super-critical fields with
next-generation laser facilities \cite{Ruf:2009zz,Dumlu:2010ua}.
Another direction for QED applications follows the lines of
\cite{Brodsky:2006in,Brodsky:2006ku}, where Fourier transform of the Deeply 
Virtual Compton Scattering
amplitude with respect to the skewness variable $\zeta$ at fixed invariant 
momentum transfer was observed to be analogous to the diffractive scattering 
of a wave in optics. In analogy with this "hadron optics", the light-front 
electron wave functions computed in \cite{Honkanen:2010rc} can be used 
to evaluate
the form factors of the electron and  thereby introduce  "electron optics".

Most importantly, the research in \cite{Vary:2009gt,Honkanen:2010rc} also 
serves as a foundation for solving other quantum field theories at strong 
coupling, such as the light-front QCD Hamiltonian in the nonperturbative 
domain.
In order to extend the research to QCD,
methods for treating the color degree of freedom in a computationally 
efficient manner  were already introduced and evaluated in \cite{Vary:2009gt} .
The next step is then to incorporate the color degree of freedom into the
quantum field theory code, that has
already passed an important accuracy test in \cite{Honkanen:2010rc}.

We will next review the method used in \cite{Vary:2009gt,Honkanen:2010rc}, 
adding some details not previously discussed.

\section{BLFQ Hamiltonian Framework}

In BLFQ approach the Hamiltonian is expressed  in terms of basis functions,
and the size of the resulting Hamiltonian matrix is regulated by imposing both
physical symmetries and
different cut-off conditions for the basis states. Increasing the size of the
basis will inevitably lead to substantial computational requirements both in
the computation of the matrix elements themselves and in the diagonalization
of the matrix. In order to be able to extrapolate the results to the continuum 
limit, where the cut-offs are removed, a rapid convergence of the results
is then highly desirable.

We define our 
light-front coordinates as $x^{\pm}=x^0 \pm x^3$, $x^{\perp}=(x^1,x^2)$,
where the variable $x^+$ is light-front time and $x^-$ is the longitudinal 
coordinate. We adopt $x^+=0$, the ``null plane", for our quantization surface.
In our choice of framework \cite{Vary:2009gt,Honkanen:2010rc} we quantize
QED on
the light-front using the light-front gauge, and add the harmonic 
oscillator potential in the transverse direction to confine the system in those
directions. To simplify the numerical work, we choose the transverse
basis function scale
and the trap scale to coincide. As a consequence, we cannot obtain zero 
external field QED results directly, but via extrapolation, as shown later.

Our basis states consist of 
2-D harmonic oscillator (HO) states, which are combined with discretized 
longitudinal  modes, plane waves satisfying selected boundary conditions. 
The HO states are
characterized by a principal quantum 
number $n$,  orbital quantum number $m$ and HO energy 
$ \Omega $. We express the 
2-D oscillator in momentum space as a 
function of the dimensionless variable  
$\rho=\vert p^\perp\vert/\sqrt{M_0\Omega}$, where 
$M_0$ has units of mass.
The orthonormalized HO wavefunctions in polar coordinates 
$(\rho,\varphi)$ are then given in terms of the 
Generalized Laguerre Polynomials, $L_n^{|m|}(\rho^2)$, by
\begin{eqnarray}
&&\Phi_{nm}(p^\perp)=\Phi_{nm}(\rho,\varphi)= \langle \rho \varphi | n m \rangle
=
\sqrt{\frac{2\pi}{M_0\Omega}}\sqrt{\frac{2n!}{(\vert m\vert+n)!}} 
e^{im\varphi}
\rho^{\vert m\vert} e^{-\rho^2/2}L^{\vert m\vert}_n(\rho^2),
\label{Eq:wfn2dHOchix}
\end{eqnarray}
with eigenvalues $E_{n,m}=(2n+|m|+1)\Omega$.
These wavefunctions are orthogonal and form a complete set of states, thus
\begin{eqnarray}
&&\sum_{nm}\Phi_{nm}^\ast(p^\perp)\Phi_{nm}(q^\perp)
={(2\pi)^2}\delta^{(2)}(p^\perp-q^\perp).
\end{eqnarray}

The longitudinal modes, $\psi_{k}$, in our basis are  defined for
$-L \le x^- \le L$ with  periodic boundary conditions (PBC) for the photon and 
antiperiodic boundary conditions (APBC) for the electron:
\begin{eqnarray}
  \psi_{k}(x^-) &=& \frac{1}{\sqrt{2L}} \, {\rm e}^{i\,\frac{\pi}{L}k\,x^-},
\label{Eq:longitudinal1}
\end{eqnarray}
where $k=1,2,3,...$  for PBC (we neglect the zero mode) and 
$k=\frac{1}{2},\frac{3}{2},\frac{5}{2},...$  for
APBC. The full 3-D single-particle basis state is defined by the product form
\begin{eqnarray}
  \Psi_{k,n,m}(x^-,\rho,\varphi) &=& \psi_{k}(x^-) \Phi_{n,m}(\rho,\varphi).
\label{Eq:totalspwfn}
\end{eqnarray}

Following Ref.\cite{Brodsky:1997de} we introduce the total invariant
 mass-squared $M^2$ for 
the low-lying physical states in terms of a Hamiltonian $H$ times a 
dimensionless integer for the total light-front momentum $K$
\begin{eqnarray}
M^2 + P_{\perp}P_{\perp} \rightarrow  M^2 + const = P^+P^- = KH
\label{Mass-squared}
\end{eqnarray}
where we absorb the constant into $M^2$.  
The non-interacting Hamiltonian $H_0$ for this system (where now
the transverse functions for both the fermion and the boson 
are taken as eigenmodes of the trap)
is then defined by the 
sum of the occupied 
modes $i$ in each many-parton state as
\begin{eqnarray}
&&  H_0 = 2M_0 P^-_c 
= \frac{2M_0\Omega}{K}\sum_i{\frac{2n_i+|m_i| +1 +
{{\bar m_i}^2}/(2M_0\Omega)
}{x_i}},
\label{Hamiltonian}
\end{eqnarray}
where ${\bar m_i}$ is the mass of the parton $i$. 
The photon mass is always set to zero in the following and the electron
mass $\bar m_e$ is set at the physical mass 0.511 MeV in our 
non-renormalized 
calculations. We also set  $M_0=\bar m_e$.

The basis is limited to fermion and 
fermion-boson states, so
the light-front QED Hamiltonian interaction terms we need are
the fermion to fermion-boson vertex, given as
\begin{equation}
V_{e\to e\gamma}=g\int dx_{+}d^2x_{\perp}
\overline{{\Psi}}(x)\gamma^\mu {\Psi}(x){A}_\mu(x)
\bigg\vert_{x^{+}=0},
\label{vertex}
\end{equation}
and the instantaneous fermion-boson interaction,
\begin{equation}
V_{e\gamma\to e\gamma}=\frac{g^2}{2}\int\!dx_+d^2x_{\!\perp}
    \ \overline{\Psi}\gamma^\mu 
    A_\mu \ \frac{\gamma^+}{i\partial^+}
    \left( \gamma^\nu  A_\nu 
    \Psi  \right)\bigg\vert_{x^{+}=0},
\label{inst}
\end{equation}
where the coupling constant $g^2=4\pi\alpha$, and $\alpha$ is the fine 
structure constant taken to be $\alpha=\frac{1}{137.036}$.
When expressing the free fermion and boson fields in terms of our basis
functions, the complete set of quantum numbers needed to specify a state 
are $\bar\alpha=(k,n,m,\lambda)$, where $\lambda$ is the helicity.
The fermion and boson annihilation operators are then written as
\begin{eqnarray}
&& b(p^\perp,\alpha)=\sum_{nm}b(\bar\alpha)\Phi_{nm}(p^\perp),\\
&& a(p^\perp,\gamma)=\sum_{nm}a(\bar\gamma)\Phi_{nm}(p^\perp),
\end{eqnarray}
where
\begin{eqnarray}
&&[a(\bar\gamma), a^\dagger(\bar\gamma^\prime)]
=\delta^{\bar\gamma^\prime}_{\bar\gamma},\\
&& \{ b(\bar\alpha), b^\dagger(\bar\alpha^\prime)\}
=\delta^{\bar\alpha^\prime}_{\bar\alpha},
\end{eqnarray}
and the truncated set of quantum numbers $\alpha=(k,\lambda)$. 
After this replacement and proper normalization, the non-spinflip
vertex terms of Eq.(\ref{vertex}) are  $\propto  M_0\Omega $, similar
to the non-interacting Hamiltonian of Eq.(\ref{Hamiltonian}),
whereas
spinflip terms are $\propto \sqrt{M_0\Omega} m_e$. 
Selecting the initial state fermion helicity in the single fermion sector
always as ``up'', the process $e\to e\gamma$ is nonzero for 3 out of 8 
helicity combinations, and the process 
 $e\gamma\to e\gamma$ is nonzero only when all 4 spin projections aligned 
(2 out of 16 combinations). The resulting Hamiltonian matrix is thus sparse.

As a symmetry constraint for the basis we fix the total angular 
momentum projection 
$J_z=M+ S=\frac{1}{2}$,
where  $M=\sum_i{m_i}$ is the total azimuthal quantum number, and  
$S=\sum_i{s_i}$
the total spin projection along the $x^-$ direction.  
For cutoffs, we select the total light-front momentum, $K$, and the maximum 
total quanta allowed in the transverse mode of each one or two-parton state, 
$N_{max}$, such that 
\begin{eqnarray}
&&\sum_i{x_i} = 1= \frac{1}{K}\sum_i{k_i},\label{momsumrule} \\
&&\sum_i{2n_i+|m_i| +1} \le N_{max},
\end{eqnarray}
where, for example, $k_i$ defines the longitudinal 
modes of Eq.(\ref{Eq:longitudinal1}) for the $i^{th}$ parton. 
Eq.(\ref{momsumrule}) signifies total light-front momentum conservation 
written in terms of boost-invariant momentum fractions, $x_i$. Since each 
parton carries at least one unit of longitudinal momentum, the basis is 
limited to $K$ partons. Furthermore, since each
parton carries at least one oscillator quanta for transverse motion, the basis 
is also limited to $N_{max}$ partons. Thus the combined limit on the number of 
partons would be min$(K,N_{max})$, if the Fock space was not truncated.

\section{BLFQ Hamiltonian results}

Here we present some numerical results from \cite{Honkanen:2010rc} for cases
where the cutoffs
for the basis space dimensions are selected such that $K$ 
increases simultaneously with the $N_{max}$. 
The resulting dimension of the Hamiltonian matrix increases rapidly. 
For $N_{max}=K=2,10,20$,
the dimensions of the corresponding symmetric $d\times d$ matrices are 
$d=2,\, 1670,\, 26990$, respectively. 
Since we employ a mix of boundary
conditions and all states have half-integer 
total $K$, we quote  $K$-values rounded downwards for convenience, except 
when the precise value is required.

\begin{figure}[h]
\includegraphics[width=0.6\textwidth]{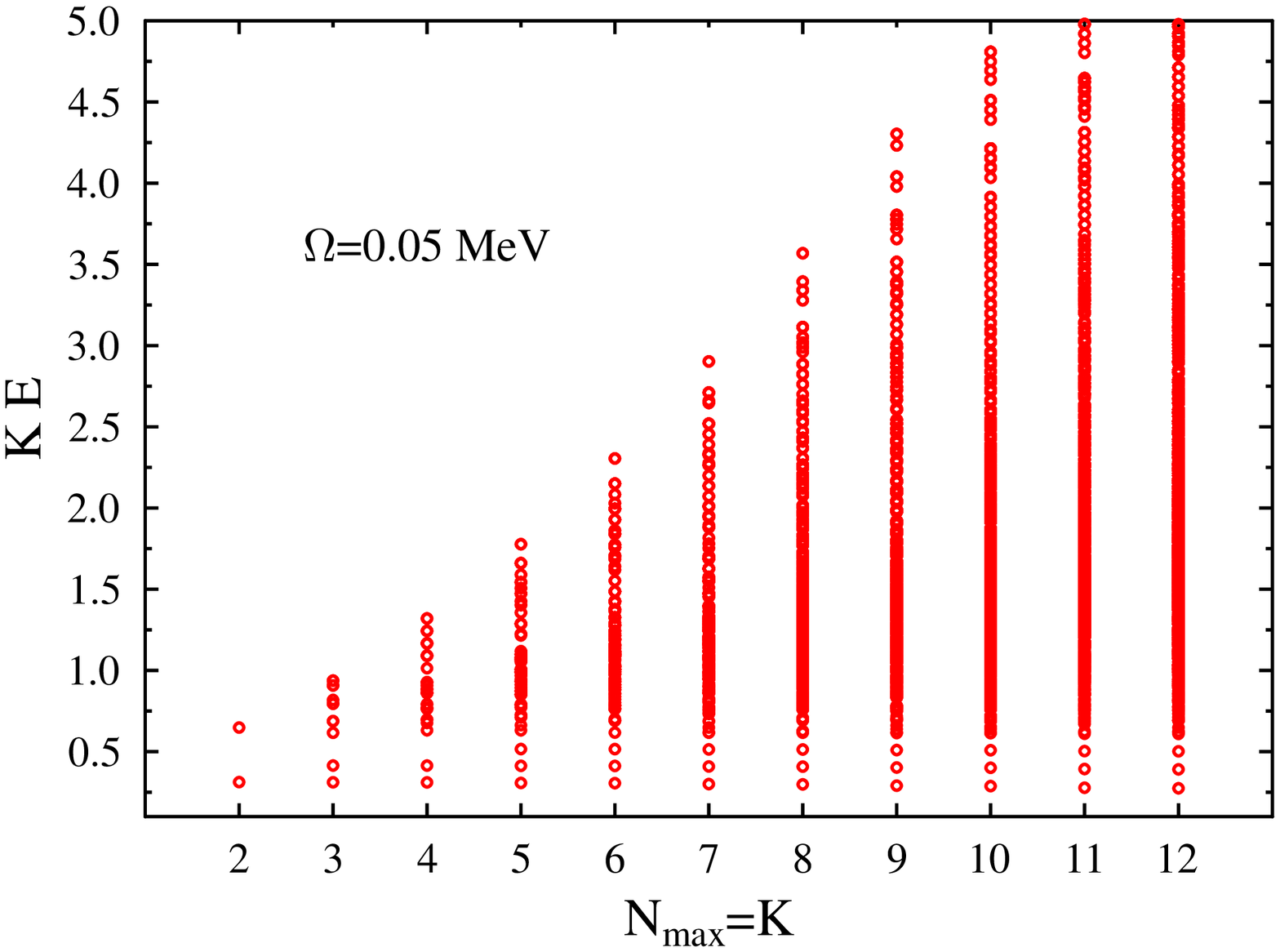}
\hspace{-1.0cm}
\includegraphics[width=0.5\textwidth]{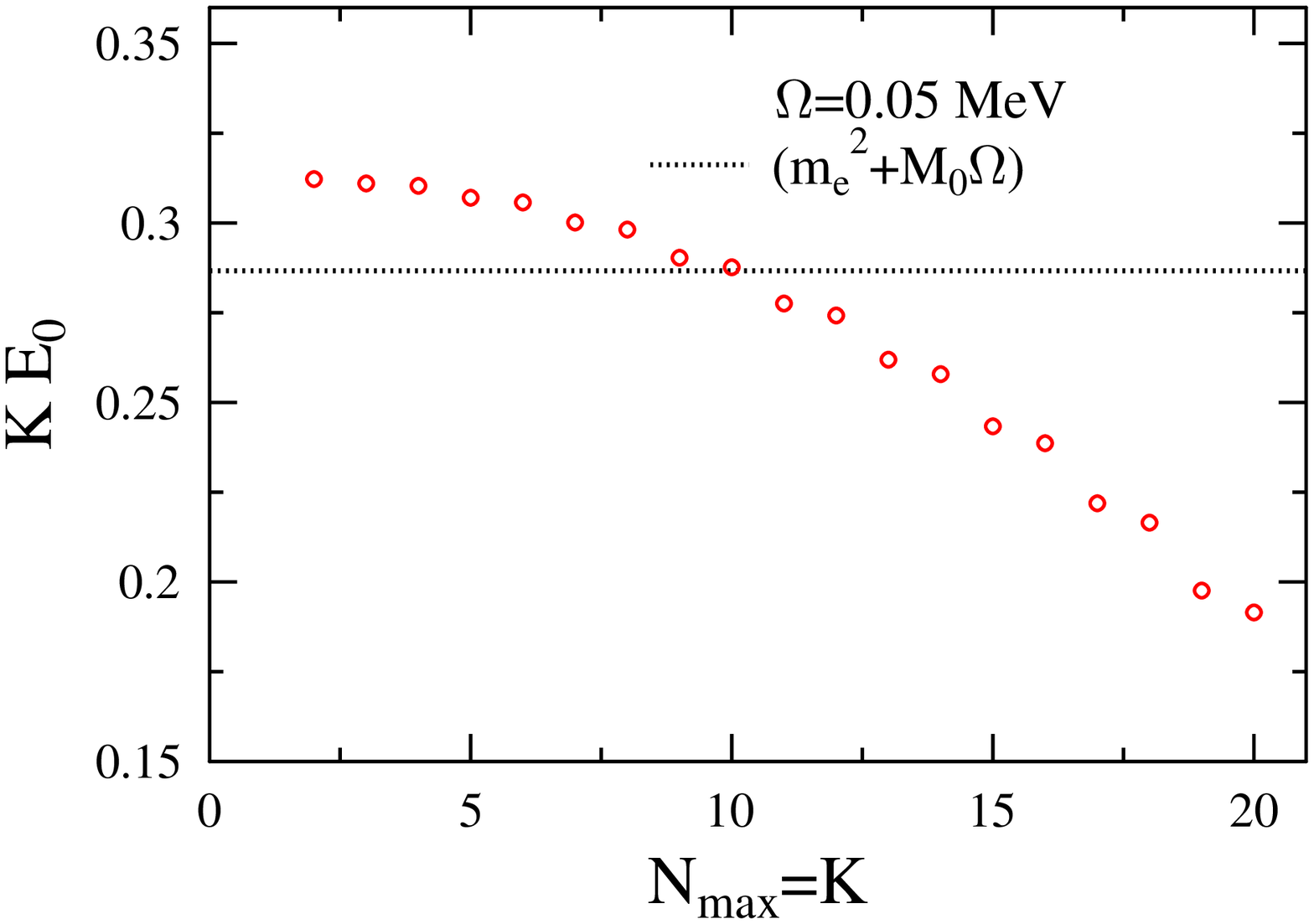}
\caption{\label{fig1} Eigenvalues (multiplied by $K$) for a 
non-renormalized 
light-front QED Hamiltonian which includes the fermion-boson vertex
and the instantaneous fermion-boson interaction without counterterms (figure
adapted from \cite{Honkanen:2010rc}).
Closeup of the lowest-lying eigenvalues on the right panel.
The basis is limited to fermion and 
fermion-boson states satisfying the symmetries. The cutoffs
for the basis space dimensions are selected such that $K$ 
increases simultaneously with the $N_{max}$.}
\end{figure}

In the left panel of Fig.\ref{fig1} we show the eigenvalues 
(multiplied by $K$) for a 
non-renormalized light-front QED Hamiltonian given in 
Eqs.(\ref{Hamiltonian},\ref{vertex},\ref{inst}), with fixed $\Omega=0.05$ MeV.
In the right panel
we show a closeup of the lowest-lying eigenvalues only.
These eigenvalues correspond to a solution dominated by the electron
with $n=m=0$, 
and are expected to be 
$K E_0={m_e^2+M_0\Omega}$, where the latter term accounts for the lowest state
of transverse
motion of the electron allowed in the chosen basis. 
In Fig.\ref{fig1} the lowest eigenvalues for a fixed 
$\Omega$ fall below that value, as the size of the matrix increases.
The contribution of the spinflip terms to the lowest eigenvalues is very small,
and, as a result, the lowest eigenvalues for a fixed  $N_{max}=K$ depend 
linearly on $\Omega$. 
Since our system is in an external field, the lowest physical mass 
eigenstate (not known experimentally) can deviate from the free-space mass.
Therefore, before renormalization, we choose only to consider cases
where the mass eigenvalue falls within 25\% of the free electron mass.

The ordering of excited states in  Fig.\ref{fig1}, due to significant 
interaction 
mixing, does not always follow the highly degenerate unperturbed spectrum of 
Eq.(\ref{Hamiltonian}).  States dominated by spin-flipped electron-photon  
components are evident in the solutions.
Nevertheless, the  lowest-lying eigenvalues appear with nearly harmonic 
separations in  Fig.\ref{fig1} as would be expected at the coupling of QED.
The multiplicity of the higher eigenstates increases rapidly with increasing
$N_{max}=K$ and the states exhibit stronger mixing with other states than
the lowest-lying states. In principle the fermion-boson basis 
states interact directly with each other in leading order through the 
instantaneous
fermion-boson interaction, but numerically the effect of this interaction
is very weak, and thus does not contribute significantly to the mixing.
Even though we work within a
Fock space approach, our numerical results should 
approximately equal the lowest
order perturbative QED results for sufficiently weak external field.
\begin{figure}[h]
\centering
\includegraphics[width=0.65\textwidth]{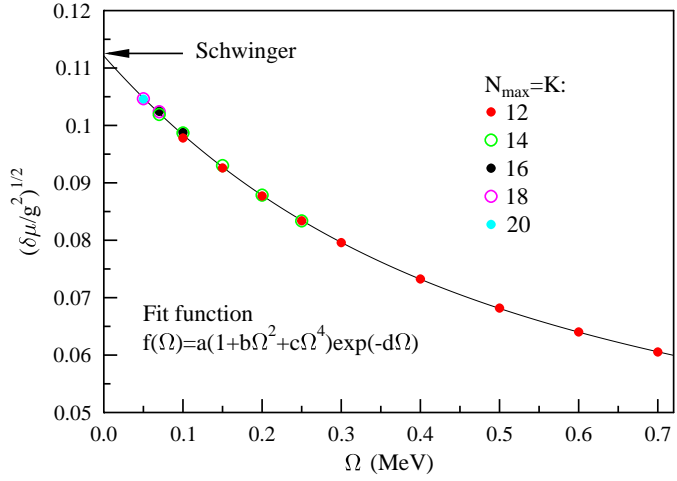}
\caption{\label{fig3} (Color online) 
Square root of the (scaled) electron anomalous magnetic 
moment as a function of the transverse external field for a sequence of 
increasing basis spaces indicated in the legend. These are non-renormalized 
results 
where the mass eigenvalue falls within 25\% of the free electron mass.
Extrapolation to zero external field yields 0.1121, compared with the 
theoretical 1-loop QED 
prediction ("Schwinger") of 0.1125.
Figure is adapted from \cite{Honkanen:2010rc}.}
\end{figure}

In Fig.\ref{fig3} we show the results for the square root of the electron 
anomalous magnetic moment (scaled),  
$\sqrt{\delta\mu / g^2}$, as a function of $\Omega$ obtained from the
lowest mass eigenstate. 
That is, we plot the magnitude of the probability 
amplitude that the
electron has its spin flipped relative to the single electron 
Fock-space component in the range where the results are converged.
For even $N_{max}=K$ the results converge 
rapidly for  $N_{max}=K\ge 14$. The results for odd cutoffs (not shown) track
even cutoff results as $N_{max}=K$ increases. Below $\Omega\lsim 0.05$ MeV 
(results not shown), in 
the weak external field region, all the interactions are quenched 
at fixed $N_{max}=K$, and not converged,  due in part to  our 
requirement that the HO basis states track the external field.
In order to compare our results with
the square root of the ratio of the Schwinger result $\frac{\alpha}{2\pi}$
for the coupling constant $g^2=4\pi\alpha$, we
perform an extrapolation of the above 
results  for $N_{max}=K=12,\dots,20$  to the
zero external field limit $\Omega=0$ Mev. 
An excellent agreement with the 
results is obtained by a fit function $f(\Omega)=
a(1+b\Omega^2+c\Omega^4)\exp(-d\Omega)$, with $a=0.1121$. This is 
$<1\%$ deviation from the Schwinger 
result of 0.1125, which is reasonable in light of our numerical accuracy and 
extrapolation uncertainties.

\begin{figure}[h]
\centering
\includegraphics[width=0.65\textwidth]{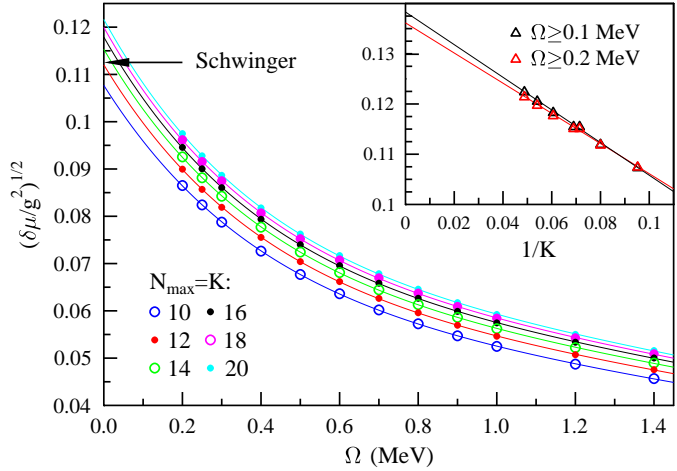}
\caption{\label{fig4} (Color online) Individual  fits to the  renormalized 
results for square root of the (scaled) electron anomalous magnetic 
moment for $N_{max}=K=10,\dots,20$.
 The inset shows
the continuum limit extrapolation of the zero external field 
results from the main panel as a function of $1/K$.
Figure is adapted from \cite{Honkanen:2010rc}.}
\end{figure}

In Fig.\ref{fig4} we renormalize our results for $\sqrt{\delta\mu / g^2}$
by applying a sector-dependent normalization scheme from 
Ref.\cite{Karmanov:2008br}.
In our present limited Fock space, we need only the mass counterterm 
$\delta m_e$. This $\delta m_e$ is added to the mass term in the
diagonal one-electron part of 
the Hamiltonian Eq.(\ref{Hamiltonian}).
In the absence of a known  experimental mass for
renormalization due to the external field, we adjust 
$\delta m_e$ such that the lowest 
eigenstate remains at $K E_0={m_e^2+M_0\Omega}$. 
To eliminate possible effects 
from the  quenched interactions at small external fields, we only include
results with the external field $\Omega\ge 0.2$ MeV. Again, individual
fits of the form given Fig.\ref{fig3} in  are 
an excellent representation of our results. The range of the extrapolated 
values is $0.1077 \le a \le 0.1216$.

The convergence with an 
increasing cutoff is now less rapid than in the non-renormalized case shown
Fig.\ref{fig3}. In order to approach the continuum limit $N_{max}=K\to\infty$,
we perform further extrapolation to the zero-$\Omega$  results
of Fig.\ref{fig4}.
The inset of Fig.\ref{fig4} shows
linear extrapolation of the results of the main 
figure in $1/K$ to the continuum
limit  $N_{max}=K\to\infty$. To verify the stability of the results, an 
extrapolation based on the  $\Omega\ge 0.1$ MeV fits (not shown)
is also given.
The extrapolated continuum values are 0.1362 (0.1383) for  
$\Omega\ge 0.2\, (0.1)$, respectively, and thus about 20\%
above the Schwinger result 0.1125. As mentioned in Sec.\ref{motivation}, 
an enhancement of this 
magnitude was also observed in related works, Ref.\cite{Brodsky:2004cx}  and 
Refs.\cite{Chabysheva:2009ez,Chabysheva:2009vm}, where one-photon truncated
light-front Hamiltonian was regulated with Pauli-Villars (PV)
regularization scheme. With PV regularization as well as in our renormalized 
results, interpreted from a perturbation theory perspective, the intermediate 
state propogators are developed from a dynamical (non-perturbative) electron 
mass rather than using the unperturbed mass needed for direct comparison with 
perturbation theory.
Thus one may appreciate why the renormalized results are distinct from the 
lowest order perturbative results.

\begin{figure}[h]
\centering
\includegraphics[width=6cm]{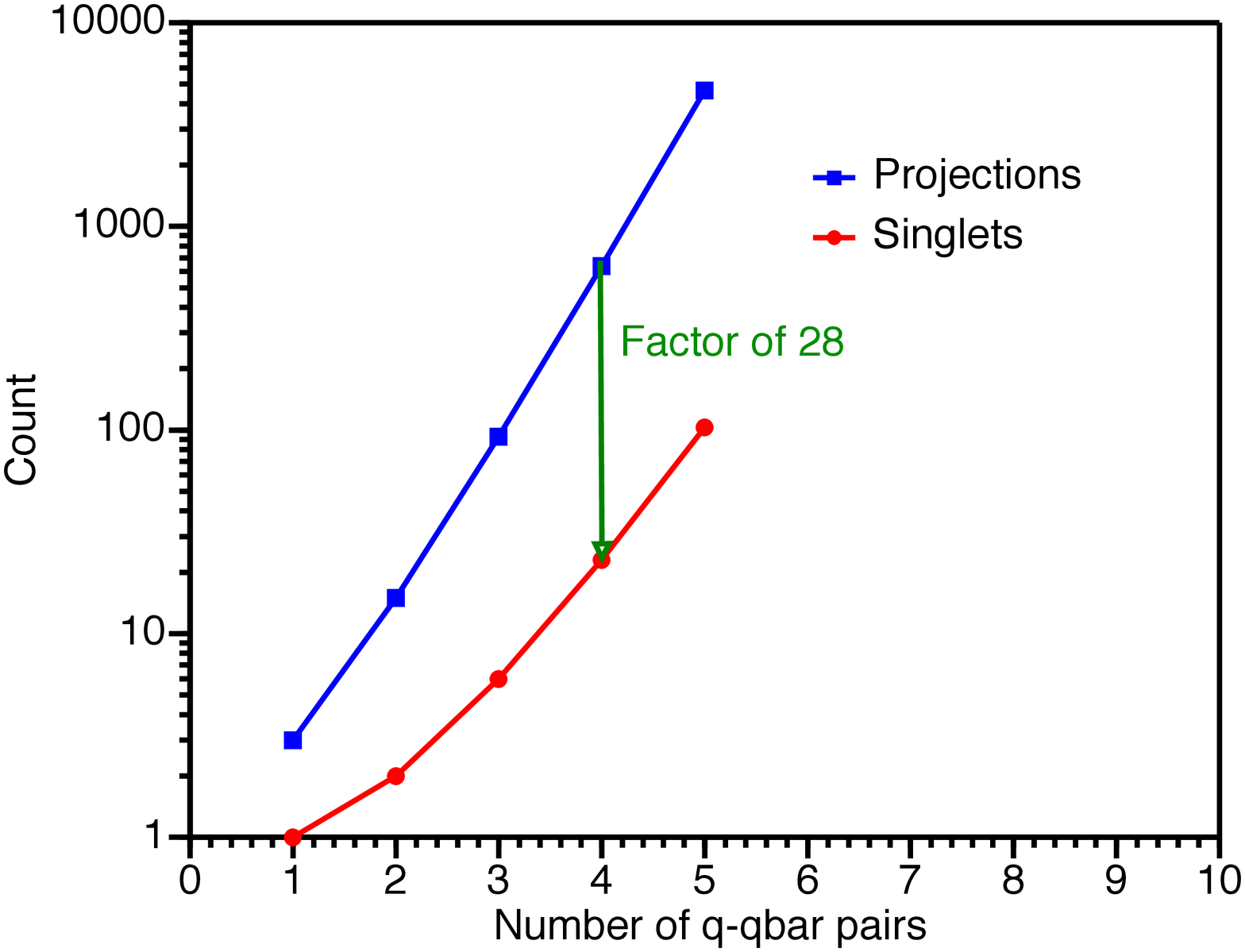}
\caption{Number of color space states that apply to
space-spin configuration of selected multi-parton states for two methods of 
enumerating the color basis states. The upper curve shows counts of all color 
configurations with zero color projection.  The lower curve counts 
global color singlets. Figure adapted Ref.\cite{Vary:2009gt}.} 
\label{colorstates}
\end{figure}

The extension of this method both to a larger Fock-space and to QCD 
will proceed as outlined in \cite{Vary:2009gt}. In the case of non-interacting
QED, the state density as a function of the state energy $E$, given by 
Eq.(\ref{Hamiltonian}), was found to increase exponentially as  $N_{max}=K$ 
increases. Similarly, implementation of the color dramatically increases the 
state density over the case of QED, as shown in Fig.\ref{colorstates}, but it 
was also found, that use of a global color-singlet constraint is effective in 
minimizing the explosion in basis space states.
We anticipate to be able to handle both of these challenges with the parallel 
codes  developed, tested and applied in
\cite{Vary:2009gt,Honkanen:2010rc},
that compute the Hamiltonian matrix and solve
for its eigenvalues and eigenvectors

\end{document}